\documentstyle[prb,aps,epsbox]{revtex}

\newcommand{\secref}[1]{\ref{#1}}

\newcommand{\Ref}[1]{Ref.~\CITE{#1}}
\newcommand{\Refs}[1]{Refs.~\CITE{#1}}
\newcommand{\figref}[1]{Fig.~\ref{#1}}

\newcommand{\ve}[1]{ {\bf #1} }

\newcommand{\lra}[1]{\left( #1 \right)}

\newcommand{\lrb}[1]{\left\{ #1 \right\}}

\newcommand{\lrc}[1]{\left[ #1 \right]}

\newcommand{\lrangle}[1]{\left<\, #1 \,\right>}

\newcommand{\lrabs}[1]{\left|\, #1 \,\right|}
\newcommand{\rbar}[2]{\left. {#1} \,\right|_{#2}}

\newcommand{\txtfrac}[2]
           {{\textstyle \frac{#1}{#2} }}

\newcommand{\Tr}{{\rm Tr}}
\newcommand{\re}{{\rm e}}
\newcommand{\ri}{{\rm i}}

\newcommand{\fdag}{ \mbox{\footnotesize \dag} }

\newcommand{\vx}{\ve{x}}
\newcommand{\vX}{\ve{X}}
\newcommand{\vk}{\ve{k}}
\newcommand{\vq}{\ve{q}}
\newcommand{\vl}{\ve{l}}
\newcommand{\vm}{\ve{m}}
\newcommand{\vkk}{\ve{k}'}
\newcommand{\vqq}{\ve{q}'}
\newcommand{\vll}{\ve{l}'}
\newcommand{\vmm}{\ve{m}'}
\newcommand{\vK}{\ve{K}}
\newcommand{\wk}{\tilde{\ve{k}}}
\newcommand{\wq}{\tilde{\ve{q}}}

\newcommand{\wkk}{\tilde{\ve{k}}'}
\newcommand{\wqq}{\tilde{\ve{q}}'}

\newcommand{\rd}{{\rm d}}
\newcommand{\del}{\partial}
\newcommand{\delt}{\partial_{t}}

\newcommand{\Del}{{\mit \Delta}}
\newcommand{\rDel}{{\rm \Delta}}

\newcommand{\cR}{{\cal R}} 

\newcommand{\cD}{{\cal D}} 
\newcommand{\cDkq}{{\cal D}_{\vk,\vq}} 

\newcommand{\mG}{{\mit \Gamma}} 
\newcommand{\mGo}{{\mit \Gamma}^{(0)}} 

\newcommand{\rI}{{\rm I}}
\newcommand{\rF}{{\rm F}}
\newcommand{\rC}{{\rm C}}

\newcommand{\tI}{t_{\rI}}
\newcommand{\tF}{t_{\rF}}
\newcommand{\intIF}{{\int_{\tI}^{\tF}\hspace{-0.2em}}} 
\newcommand{\intIt}{{\int_{\tI}^{t}\hspace{-0.2em}}} 
\newcommand{\intIs}{{\int_{\tI}^{s}\hspace{-0.2em}}} 

\newcommand{\psio}{\hat{\psi}}
\newcommand{\psid}{\hat{\psi}^{\fdag}}
\newcommand{\psiok}{\hat{\psi}_{\vk}}

\newcommand{\psidq}{\hat{\psi}^{\fdag}_{\vq}}

\newcommand{\psiD}{\psi_{\rDel}}
\newcommand{\psiC}{\psi_{\rC}}
\newcommand{\psidD}{\psi^{\ast}_{\rDel}}
\newcommand{\psidC}{\psi^{\ast}_{\rC}}
\newcommand{\psikD}{\psi_{\vk,\rDel}}
\newcommand{\psikC}{\psi_{\vk,\rC}}
\newcommand{\psiqD}{\psi_{\vq,\rDel}}
\newcommand{\psiqC}{\psi_{\vq,\rC}}
\newcommand{\psidkD}{\psi^{\ast}_{\vk,\rDel}}
\newcommand{\psidkC}{\psi^{\ast}_{\vk,\rC}}

\newcommand{\epk}{\epsilon_{\vk}}
\newcommand{\epq}{\epsilon_{\vq}}

\newcommand{\omk}{\omega_{\vk}}
\newcommand{\omq}{\omega_{\vq}}

\newcommand{\ommk}{\omega_{\vm-\vk}}

\newcommand{\oml}{\omega_{\vl}}
\newcommand{\omml}{\omega_{\vm-\vl}}
\newcommand{\omqklm}{\omega^{(2)}_{\vq,\vk,\vl,\vm}}
\newcommand{\omkqlm}{\omega^{(2)}_{\vk,\vq,\vl,\vm}}

\newcommand{\Jkq}{J_{\vk,\vq}}
\newcommand{\JD}{J_{\rDel}} 
\newcommand{\JC}{J_{\rC}} 
\newcommand{\JkqD}{J_{\vk\vq,\rDel}} 
\newcommand{\JkqC}{J_{\vk\vq,\rC}} 
\newcommand{\Jo}{J^{(0)}}
\newcommand{\Jokq}{J^{(0)}_{\vk,\vq}}
\newcommand{\JoD}{J^{(0)}_{\rDel}}
\newcommand{\JoC}{J^{(0)}_{\rC}}
\newcommand{\JokqD}{J^{(0)}_{\vk\vq,\rDel}}
\newcommand{\JokqC}{J^{(0)}_{\vk\vq,\rC}}
\newcommand{\DelJ}{\Del J}
\newcommand{\DelJkq}{\Del J_{\vk,\vq}}
\newcommand{\DelJD}{\Del J_{\rDel}}
\newcommand{\DelJC}{\Del J_{\rC}}

\newcommand{\ID}{I_{\rDel}}
\newcommand{\deltakq}{\delta_{\vk,\vq}}

\newcommand{\Go}{G^{(0)}}

\newcommand{\Gokq}{G^{(0)}_{\vk, \vq}}

\newcommand{\Gotil}{\tilde{G}^{(0)}}

\newcommand{\gR}{g^{\rm R}}
\newcommand{\gA}{g^{\rm A}}
\newcommand{\gC}{g^{\rC}}
\newcommand{\gRkq}{g^{\rm R}_{\vk,\vq}}
\newcommand{\gAkq}{g^{\rm A}_{\vk,\vq}}
\newcommand{\gCkq}{g^{\rC}_{\vk,\vq}}

\newcommand{\gRtil}{\tilde{g}^{\rm R}}
\newcommand{\gAtil}{\tilde{g}^{\rm A}}
\newcommand{\gCtil}{\tilde{g}^{\rC}}
\newcommand{\gDtil}{\tilde{g}^{\rDel}}

\newcommand{\zlL}{z_{\vl, \vll}}
\newcommand{\zkq}{z_{\vk, \vq}}
\newcommand{\zkQ}{z_{\vk, \vqq}}

\newcommand{\zmqmQ}{z_{\vm-\vq, \vm-\vqq}}

\newcommand{\zmlML}{z_{\vm-\vl, \vmm-\vll}}

\newcommand{\zo}{z^{(0)}}

\newcommand{\zokq}{z^{(0)}_{\vk,\vq}}
\newcommand{\zoqk}{z^{(0)}_{\vq,\vk}}

\newcommand{\zdqQ}{z^{\ast}_{\vq, \vqq}}

\newcommand{\zdqk}{z^{\ast}_{\vq, \vk}}

\newcommand{\zdmkmQ}{z^{\ast}_{\vm-\vk, \vm-\vqq}}
\newcommand{\zdmkMQ}{z^{\ast}_{\vm-\vk, \vmm-\vqq}}

\newcommand{\zbarlL}{\bar{z}_{\vl, \vll}}

\newcommand{\zbarmlML}{\bar{z}_{\vm-\vl, \vmm-\vll}}

\newcommand{\zbardqQ}{\bar{z}^{\ast}_{\vq, \vqq}}

\newcommand{\zbardmkMQ}{\bar{z}^{\ast}_{\vm-\vk, \vmm-\vqq}}

\newcommand{\zbarokq}{\bar{z}^{(0)}_{\vk, \vq}}
\newcommand{\zbaroqk}{\bar{z}^{(0)}_{\vq, \vk}}

\newcommand{\zC}{z_{\rC}}
\newcommand{\zkqC}{z_{\vk\vq,\rC}}

\newcommand{\zD}{z_{\rDel}}
\newcommand{\zkqD}{z_{\vk\vq,\rDel}}

\newcommand{\ZZqklm}{Z^{(2)}_{\vq,\vk,\vl,\vm}}
\newcommand{\ZZdkqlm}{Z^{(2)\ast}_{\vk,\vq,\vl,\vm}}

\begin{document}
\draft
\title{ Kinetic-Theoretic Description 
        based on Closed-Time-Path Formalism
       }
%
\author{Jun Koide}
\address{Department of Physics, 
         Faculty of Science and Technology, 
         Keio University, Yokohama 223-8522, japan
        }
\date{\today}
\maketitle
\begin{abstract}
Utilizing a non-equilibrium Green function 
like the generalized Kadanoff-Baym ansatz,
a systematic perturbative method is presented
to calculate the expectation value of an arbitrary physical quantity 
under the restriction that the Wigner distribution function is fixed.
It is shown that,
in the diagrammatic expression of the quantity,
a certain part of contributions can be eliminated 
due to the restriction.
Together with the quantum kinetic equation,
this method provides a basis for the kinetic-theoretical description.
\end{abstract}
%
%
\section{Introduction}
\label{intro}
The non-equilibrium state of a dilute gas system 
is considered to be described 
by the one-particle distribution function (1PDF),
and such an approach to the non-equilibrium system
is called the `kinetic theory'~\cite{Zubarev}.
In the kinetic theory,
the 1PDF is treated as the independent dynamical variable of the system,
and all the physical quantities are determined by the 1PDF.
The kinetic equation is an equation of motion of the 1PDF, 
and the dynamics in the kinetic theory is described by this equation.

A lot of works has been done on the derivation 
of the quantum kinetic equation (QKE) 
in the framework of the Green function technique.
Perhaps,
the most popular approach is the generalized Kadanoff-Baym (GKB) 
formalism~\cite{KB,Lipavsky}
which utilizes an ansatz for the non-equilibrium Green function
called the GKB ansatz.
The GKB ansatz can be expressed 
by the usual equilibrium Keldysh Green function 
in which the equilibrium 1PDF is replaced by a non-equilibrium one.

An alternative approach is the counter-term method 
based on the CTP formalism~\cite{Lawrie,Niegawa},
or on the thermo-field dynamics~\cite{Umezawa}.
In this approach,
a counter-term,
in which the non-equilibrium properties are included,
is first introduced into the CTP or thermo-field Lagrangian,
and then the unperturbed propagator gets the similar structure 
as the GKB ansatz.
 
Although the QKE can be derived in these approaches,
the kinetic theory is not completely constructed
because they do not give a proper method to express
the expectation values of physical quantities in terms of the 1PDF.
Of course,
the expectation value of a physical quantity of interest 
can be calculated perturbatively 
by the usage of the non-equilibrium propagators mentioned above,
and it becomes a functional of the 1PDF.
But if we want to obtain a functional of the 1PDF,
the value of the 1PDF must be fixed from the exterior,
and the integrations over the microscopic fields should be carried out 
under the restriction due to the fixing of the 1PDF.
This restriction has not been considered 
in the above formalisms,
and hence they do not give a complete basis for the kinetic theory.

In this paper,
we present a systematic perturbative method 
to calculate the expectation value of any physical quantity
as a functional of the Wigner distribution function (WDF),
which plays the role of the 1PDF in quantum theory.
Our approach,
is somewhat different from the GKB or the counter-term method.
It is based on the inversion method~\cite{Fukuda,suppl}.
An inversion-method approach to derive the QKE 
was presented in \Refs{Koide,Koide2,Koide3,thesis},
and the problem of calculating the physical quantities 
in terms of the WDF is partly solved there;
We first introduce an external source $J$ to probe the WDF $z$,
and calculate the physical quantity
as a functional of the source $Q[J]$.
Then we express the source as a functional of the WDF
($J=J[z]$ which is an inversion of $z=z[J]$),
and by substituting it into the above calculated $Q[J]$,
we can write the quantity by the WDF.
In this approach,
the propagator is a functional of the external source $J$,
and the perturbative calculation can be done 
without the restriction of fixing the WDF.
The calculations in our formulation
indeed generates different results 
from those obtained by the perturbative calculation using,
e.g.,
the GKB ansatz.
By the substitution of the inverted relation,
some contributions from the diagrams, 
which will be present in the calculations with other formalisms,
are canceled.

We show that the contributions which are canceled 
can be expressed by a corresponding time-ordered diagrams:
The contributions from a diagram in the non-equilibrium theory
can be classified by the temporal order of the vertices in the diagram,
and to each temporal ordering of the vertices,
a time-ordered diagram (called a configuration in the article)
corresponds.
Then if an obtained configuration
can be separated into two parts 
by cutting two propagators at the same instant,
the contribution from that configuration is canceled.
Because the propagator used in our method 
has the form of the GKB ansatz,
this will also provide a basis for the kinetic theoretic description 
in the GKB formalism.

In the course of proving this property,
we reformulate the inversion method approach
in the framework of the Legendre transformation~\cite{suppl,CTP,Chou}.
The definition of non-equilibrium generating functional 
is slightly modified 
in a way characteristic to the non-equilibrium theory,
and the effective action is defined 
as the Legendre transformation of it.
Then the diagrammatic rule for the effective action
discussed in \Ref{Yokojima} can be utilized 
with an extension to the non-equilibrium case.
By virtue of this,
the QKE can also be expressed in a compact form
which is finally given as (\ref{QKE}).

In the next section,
we summarize the inversion method approach to the QKE,
and reformulate it in the terminology of Legendre transformation.
Then in \secref{diagrammer},
the diagrammatic rule for the kinetic theory is discussed:
The rule to calculate the expectation value as a functional of the WDF
is presented in \secref{Q-rule},
and the rule to derive the QKE is in \secref{QKE-rule}.

\section{Inversion method approach to the kinetic equation}
\label{inversion}

In this section,
we describe the inversion method approach 
to the QKE~\cite{Koide,Koide2,Koide3}.

\subsection{Probing source and the Green function}
\label{source-inhom}
The system to be considered is the same as in \Ref{Koide3};
a non-relativistic bosonic field described by the Hamiltonian
$
 \hat{H} = \hat{H}_{0}+\hat{H}_{\rm int}
$
with 
\begin{equation}
  \hat{H}_{0}  
  = \sum_{\vk} \epk \psio^{\fdag}_{\vk} \psio_{\vk} 
,
\label{H0}
\hspace{2em}
  \hat{H}_{\rm int}  
  = \frac{\lambda}{4}\sum_{\vk,\vkk,\vq} 
    \psio^{\fdag}_{\vk+\vq} \psio^{\fdag}_{\vkk-\vq} 
    \psio_{\vk} \psio_{\vkk} 
,
\label{Hint}
\end{equation}
and a spatially inhomogeneous initial density matrix $\hat{\rho}$.
We consider the case 
that the interaction $\hat{H}_{\rm int}$ can be treated perturbatively,
and for simplicity,
the initial correlation is not taken into account. 
See \Ref{Koide2} for the treatment of the initial correlation.

In quantum statistical physics,
the natural alternative of the 1PDF 
will be the Wigner distribution function (WDF) defined as
\begin{equation}
 f_{\vK}(\vX,t)
  = \int\frac{\rd\Del\vx}{V} \,\re^{-\ri\vK\cdot\Del\vx}
    \lrangle{ \psid(\vX-\frac{\Del\vx}{2},t)
              \psio(\vX+\frac{\Del\vx}{2},t)
            }
,
\label{Wigner}
\end{equation}
where
$  
\psio(\vx) 
 = \frac{1}{\sqrt{V}} \sum_{\vk} \re^{\ri\vk\cdot\vx}\psio_{\vk} 
$
and the angular bracket implies the average 
over initial density matrix $\hat{\rho}$;
$\lrangle{\cdots}=\Tr\hat{\rho}\cdots$.
As in \Ref{Koide3},
for the sake of perturbative calculation,
it is more convenient to work with the Fourier transform of the WDF
defined as
\begin{equation}
  \zkq(t)
  \equiv 
     \lrangle{ \psidq(t) \psiok(t) }
  =  \int \rd \vx \: 
     \re^{-\ri (\vk-\vq) \cdot \vx} f_{\frac{\vk+\vq}{2}}(\vx,t)
,
\label{defz}
\label{n2z}
\end{equation}
to which we refer simply as the WDF in the following.
Note that $\zdqk=\zkq$ holds 
due to the hermitian property of $\hat{\rho}$.

Within the CTP formalism,
Eq.~(\ref{defz}) can be represented as
\begin{equation}
  \zkq(t)
  \propto
        \int\lrc{ \rd \psi_{1} \rd \psi_{2} }
        \psi^{\ast}_{\vq}(t) \psi_{\vk}(t) 
        \re^{ \frac{\ri}{\hbar}\intIt\rd s
              \lra{ L(\psi_{1})-L(\psi_{2}) }
            }
        \lrangle{ \psi_{1,\rI} \lrabs{ \hat{\rho} } \psi_{2,\rI} }
.
\label{z}
\end{equation}
In the inversion method approach to the kinetic theory,
we introduce a probing source $J$ for $z$,
and calculate $z[J]$ as a functional of the source.
By inverting the relation as $J=J[z]$,
the QKE is obtained as an equation of motion for $z$ by setting $J=0$.
According to \Ref{Koide},
the proper way to introduce the source $\Jkq$ is 
that the source is built into the quadratic form 
of the free part of the Lagrangian in (\ref{z})
$
 L_{0}(\psi_{1})-L_{0}(\psi_{2}) 
  = \sum_{\vk\vq,ij}\psi^{\ast}_{\vk,i} \cD_{\vk\vq,ij}\psi_{\vq,j}
$ 
by
\begin{eqnarray}
  \cDkq
  &=& \lra{\begin{array}{cc}
            (\ri \hbar \delt -\epk) \deltakq +\ri \Jkq(t)
           & -\ri \Jkq(t)
           \\
             -\ri \Jkq(t)
           & -(\ri \hbar \delt -\epk) \deltakq +\ri \Jkq(t)
           \\
           \end{array}
          }
.
\nonumber\\
\label{D1-inhom}
\end{eqnarray}
An inverse of this matrix leads to the $2\times2$-Green function,
which is a functional of the source $J$.
From the relation $\cD\Go=-\ri\hbar$,
we get
\begin{equation}
 \Go_{\vk,\vq}[t,s;J]
   = -\theta(t-s)\re^{-\ri\omk (t-s)} 
      \lra{\begin{array}{cc}
             \zbarokq(s) & \zokq(s)\\
             \zbarokq(s) & \zokq(s)\\
           \end{array}
          }
     -\theta(s-t)\re^{ \ri\omq (s-t)} 
      \lra{\begin{array}{cc}
             \zoqk(t)    & \zoqk(t)    \\
             \zbaroqk(t) & \zbaroqk(t) \\
           \end{array}
          }
,
\label{Go0}
\end{equation}
where $\zokq$ is an unperturbed WDF as a functional of $J$ given by
\begin{equation}
  \zokq\lrc{t;J} 
   =  \re^{-\ri(\omk-\omq)(t-s)} \zokq(\tI)
     +\frac{1}{\hbar}\intIt\rd s\, \re^{-\ri(\omk-\omq)(t-s)} \Jkq(s)
,
\label{z0}
\end{equation}
and we have used 
$\zbarokq = \zokq+\delta_{\vk,\vq}$ and $\omk=\epk/\hbar$.
The unperturbed WDF satisfies an equation of motion
\begin{equation}
  \Jkq(t) = \lrb{\hbar \del_{t}+\ri(\epk-\epq)} \zokq(t) 
,
\label{J0}
\end{equation}
and if we replace $\zo$ by $z$ in (\ref{J0}),
this gives a functional expression of the source $J$ in terms of $z$
in the lowest order of the perturbative inversion.
We denote it as $\Jo[z]$.
Note that,
whenever the source has an index of the perturbative order as $J^{(i)}$,
it should be understand as a functional of $z$.
By the inversion method,
we can calculate a perturbative correction $\DelJ[z]$ to (\ref{J0}),
and up to the second order of the perturbation,
the source is expressed by $z$ as
\begin{eqnarray}
 \Jkq(t)
 &=& \Jokq[z] +\DelJkq[z] 
\nonumber\\
 &=& \lra{ \hbar \del_{t} +\ri \lra{\epk-\epq} }\zkq(t) 
    +\ri\lambda \sum_{\vqq,\vm} 
      \lrb{ \zdqQ \zdmkmQ -\zkQ \zmqmQ }(t)
\nonumber\\
 & &-\frac{\lambda^{2}}{2\hbar}\sum_{\vl,\vm}
     \intIs \rd s 
     \lrb{ \re^{ \ri\omqklm(t-s)} \ZZqklm(s) 
          +\re^{-\ri\omkqlm(t-s)} \ZZdkqlm(s) }
,
\label{J2}
\end{eqnarray}
where
$
  \omqklm 
   = \omq+\ommk-\oml-\omml
$
and
\begin{equation}
  \ZZqklm
   =  \sum_{\vqq,\vll,\vmm}
      \lrb{ \zbardqQ \zbardmkMQ \zlL \zmlML 
           -\zdqQ \zdmkMQ \zbarlL \zbarmlML }
.
\label{ZZdef}
\end{equation}
A QKE follows from (\ref{J2}) after the removal of the source.
This QKE is reduced to the usual Boltzmann equation 
after the Markovian and local approximations~\cite{Koide3}.

\subsection{kinetic theoretic description in the inversion method}
\label{problem}

In the kinetic theory,
the 1PDF is considered to be an independent dynamical variable
and the kinetic equation describes its dynamics.
This means a coarse graining 
from the microscopic field variables to the 1PDF.
All other quantities should be expressed in terms of the 1PDF,
and their dynamics should follow from the kinetic equation.
Since such physical quantities may be defined microscopically 
by a temporary-local functions of the field variables as 
$\hat{Q}=Q(\hat{\psi})$,
in order to obtain a complete framework of kinetic theory,
we must express their expectation values as functionals of the 1PDF.

In the inversion method approach,
this can be realized 
by first calculating perturbatively the expectation value 
$Q(t)=\langle \hat{Q}(t) \rangle$
with the use of the propagator $-\Go[J]$ in previous subsection,
and then by substituting the source written by the WDF as in (\ref{J2})
into the obtained functional.
The former procedure will provide us 
with the expectation value as a functional of the source $J$,
and the latter reduces it into a functional of the WDF $z$.

From some explicit calculations~\cite{thesis},
we can see the following fact:
After the substitution of $J=\Jo[z]+\DelJ[z]$ in the second step,
if we expand the obtained expression around $\Jo[z]$,
the contributions due to the perturbative correction $\DelJ[z]$
cancels some part of the unperturbed contributions.
Such an expansion can be expressed diagrammatically
by the usage of a propagator $-\Go[J=\Jo[z]]$,
and the above mentioned cancellation 
implies that some part of the diagram should be omitted 
if we want to evaluate the expectation value 
as a functional of the WDF.

From this observation,
it is expected that
a simplified procedure to calculate the expectation value $Q[z]$ exists.
Our problem in this paper is to find 
what kind of the diagram should be retained 
in the evaluation of the $Q[z]$
if the diagrams are written with the propagator $-\Go[J=\Jo]$.
In fact,
the propagator $-\Go[J=\Jo[z]]$ has the form of the GKB ansatz 
with the free particle approximation of the spectral function;
it can be obtained by replacing $\zo[J]$ in (\ref{Go0}) by $z$.
So our consideration here 
also provides the way to calculate $Q[z]$ in the GKB formalism.

\subsection{Formulation with Legendre transformation}

In order to discuss the problem settled in the previous subsection,
it is convenient to rewrite the inversion method 
in the framework of the Legendre transformation 
using the `physical representation' of the CTP formalism.
The physical representation of the CTP formalism 
is introduced by a simple transformation of the variables
from $\psi_{1}$ and $\psi_{2}$ to $\psiC$ and $\psiD$,
which is defined as
\begin{equation}
  \psiC=\frac{\psi_{1}+\psi_{2}}{2},
\hspace{2em}
  \psiD=\psi_{1}-\psi_{2}.
\end{equation}
Then the free part of the Lagrangian is rewritten as
\begin{equation}
  L_{0}(\psiC,\psiD) 
   =  \sum_{\vk,\vq}
      \begin{array}{c}
        \lra{\psidkC \psidkD} \\
        {}                    \\
      \end{array}
      \lra{\begin{array}{cc}
             0               & (\ri \hbar \delt -\epk) \deltakq \\
             (\ri \hbar \delt -\epk) \deltakq &     \ri \JkqC(t) \\
           \end{array}
          }
      \lra{\begin{array}{c}
             \psiqC \\
             \psiqD \\
           \end{array}
          }
,
\label{L0CD}
\end{equation}
where we have denoted the source $J$ in (\ref{D1-inhom}) as $\JC$
for convenience.
We can see the source $\JC$ is simply coupled to $\psidD\psiD$.
Correspondingly,
the Green function $\Go[\JC]$ becomes
\begin{equation}
  \Gokq[t,s;\JC]
   =  \lra{\begin{array}{cc}
             \gCkq[t,s;\JC] & \gRkq(t,s) \\
             \gAkq(t,s)     & 0          \\
           \end{array}
          }
,
\label{Go}
\end{equation}
where the respective components are defined as
\begin{eqnarray}
  \gCkq[t,s;\JC] 
  &=&-\theta(t-s)\re^{-\ri\omk (t-s)} \lra{ \zokq[s;\JC]+\frac{1}{2} }
     -\theta(s-t)\re^{ \ri\omq (s-t)} \lra{ \zoqk[t;\JC]+\frac{1}{2} }
,
\label{gC}
\\
  \gRkq(t,s)
  &=&-\theta(t-s)\re^{-\ri\omk (t-s)} \delta_{\vk,\vq}
,
\label{gR}
\\
  \gAkq(t,s)
  &=& \theta(s-t)\re^{ \ri\omq (s-t)} \delta_{\vq,\vk}
.
\label{gA}
\end{eqnarray}

To use the Legendre transformation formalism,
we must introduce another source $\JD$ coupled to $\psidC\psiC$,
and define the generating functional $W$ as
\begin{eqnarray}
  \re^{\frac{\ri}{\hbar}W[\JC,\JD,\ID]}
  &\equiv&
      \int\lrc{\rd\psiC\rd\psiD}
      \lrangle{\psi_{\rC,\rI}+\txtfrac{1}{2}\psi_{\rDel,\rI} 
               \lrabs{ \hat{\rho} }
               \psi_{\rC,\rI}-\txtfrac{1}{2}\psi_{\rDel,\rI} 
              }
\nonumber\\
   & &\times
      \re^{\frac{\ri}{\hbar}\intIF \rd t
           \lrb{ L_{0}(\psiC,\psiD)-V(\psiD,\psiC)
                +\sum_{\vk,\vq} \JkqD \psidkC\psiqC
               }
          }
\nonumber\\
   & &\times
      \re^{\frac{\ri}{\hbar}\intIF \rd t \,\ID Q(\psiC)}
,
\label{W0}
\end{eqnarray}
where $V$ is the interaction part of the CTP Lagrangian 
$H_{\rm int}(\psi_{1})-H_{\rm int}(\psi_{2})$
written in terms of $\psiC$ and $\psiD$.
The source $\JD$ is unphysical
in the sense that the expectation value of an hermitian operator
is not guaranteed to be real under the existence of this source.
It is just introduced 
so that we can write the WDF $z$ 
by a derivative of the generating functional,
and should be removed after all the calculation.
For the same reason,
we have introduced the source $\ID$ which is coupled to $Q(\psiC)$;
replacement of $\psio$ by $\psiC$ 
in the time-local composite operator $Q(\psio)$ 
in which we are interested.
Note that all the integrands in the exponent of (\ref{W0}) 
are local in time.

Here we define two variables
\begin{equation}
  \zkqC(t)
  \equiv  \frac{\delta W[\JD,\JC,\ID]}{\delta \JkqD(t)}
,
\hspace{1em}
  \zkqD(t)
  \equiv  \frac{\delta W[\JD,\JC,\ID]}{\delta \JkqC(t)}
.
\label{zdef}
\end{equation}
When the sources are removed,
$\zC$ is reduced to 
$
  \langle T\psid\psio+\psio\psid
         +\psid\psio+\tilde{T}\psid\psio 
  \rangle
  = \zkq+\delta_{\vk,\vq}/2
$,
and $\zD$ is to
$
  \ri\langle T\psid\psio-\psio\psid
            -\psid\psio+\tilde{T}\psid\psio 
     \rangle
   = 0
$.
Note that we regard $\psi^{\ast}\psi$ as $\psi^{\ast}(t+0)\psi(t)$
in the course of the path integration.
Particularly,
$\zD=0$ is realized by removing only the unphysical source $\JD$,
and in this case,
$\zC$ becomes a functional of $\JC$ as $\zkq[\JC]+\delta_{\vk,\vq}/2$.
Of course,
the non-equilibrium expectation value of symmetrized $\hat{Q}$ 
is obtained as a functional of the sources $\JC$ by
\begin{equation}
  Q[t;\JC]
   = \rbar{\frac{\delta W[\JD,\JC,\ID]}{\delta \ID(t)}}
          {\JD=\ID=0}
.
\label{Q1}
\end{equation}

To use the variables $\zC$ and $\zD$ as the independent variables,
we define the Legendre transformation of $W$ by
\begin{equation}
  \mG[\zC,\zD;\ID]
  \equiv W[\JD,\JC,\ID] 
        -\sum_{\vk,\vq}\intIF\rd t \lra{\JkqD\zkqC +\JkqC\zkqD}
,
\label{Gamma0}
\end{equation}
where $\JD$ and $\JC$ are functional of $\zC$ and $\zD$,
which are obtained by solving (\ref{zdef}).
Now we can obtain the expectation value of $\hat{Q}$
as a functional of the WDF $\zC$ by 
\begin{equation}
  Q[t;\zC]
   = \rbar{\frac{\delta \mG[\zC,\zD;\ID]}{\delta \ID(t)}}
          {\zD=\ID=0}
.
\label{Q2}
\end{equation}
Moreover,
from an identity of the Legendre transformation,
we have
\begin{equation}
  \JkqC(t)
  = -\frac{\delta \mG[\zC,\zD;\ID]}{\delta \zkqD(t)}
,
\hspace{1em}
  \JkqD(t)
  = -\frac{\delta \mG[\zC,\zD;\ID]}{\delta \zkqC(t)}
.
\label{Jderiv}
\end{equation}
If we remove the unphysical source $\JD$,
the first equation become an equation of motion 
of $\zkqC=\zkq[\JC]+\delta_{\vk,\vq}/2$ which corresponds to (\ref{J2}),
and finally the removal of $\JC$ 
reduces the equation of motion to the QKE.
In this sense,
$\mG$ is referred to as the effective action.

\section{Diagrammatic rule for kinetic theory}
\label{diagrammer}

Diagrammatic expression of the effective action $\mG$ 
is well investigated.
For an expectation value of non-local product 
$\langle \psid(t)\psio(s) \rangle$,
the effective action is expressed simply 
by the two particle irreducible (2PI) diagrams~\cite{DM}.
Here,
2PI diagram is a diagram which cannot be separated 
by cutting any pair of propagators.
For the expectation value of a local product,
such as the 1PDF,
the situation is more complicated.
In this subsection,
we utilize the rules presented in \Ref{Yokojima} 
with a non-equilibrium extension,
and clarify the meaning of the rule.
By use of the rule,
the QKE can also be rewritten in a compact form.
For notational simplicity,
the time arguments and wave-number indices 
will not explicitly be written if it is not misleading.

\subsection{Diagrammatic expression of the effective action}

First we consider a diagrammatic expansion
of the generating functional $W$.
The building blocks of the diagram 
are a $2\times2$-propagator $-\Gotil$ given below,
an interaction vertex $V(\psiD,\psiC)$ given in (\ref{W0})
and an external leg $Q(\psiC)$ coupled to $\ID$.
In the diagram,
an arrow expresses the contraction operator
\begin{equation}
  -\sum_{\vk,\vq}\int\rd t \rd s
      \begin{array}{c}
        \lra{\frac{\delta}{\delta\psikC(t)} \: 
             \frac{\delta}{\delta\psikD(t)}
             }                                 \\
        {}                                     \\
      \end{array}
      \Gotil_{\vk,\vq}(t,s)
      \lra{\begin{array}{c}
             \frac{\delta}{\delta\psi_{\vq,\rC}^{\ast}(s)}  \\
             \frac{\delta}{\delta\psi_{\vq,\rDel}^{\ast}(s)}\\
           \end{array}
          }
.
\end{equation}
The $2\times2$-Green function $\Gotil$ 
is defined as an inverse of the matrix 
in bilinear form of the exponent in (\ref{W0});
\begin{equation}
  \Gotil[\JC,\JD]
   \equiv
     \frac{\ri}{\hbar}
     \lra{\begin{array}{cc}
            \JD                       & \ri \hbar \delt -\epsilon \\
            \ri \hbar \delt -\epsilon & \ri \JC                   \\
          \end{array}
         }^{-1}
   = \lra{ \begin{array}{cc}
             \gCtil & \gRtil \\
             \gAtil & \gDtil \\
           \end{array}
         }
,
\label{Gotil}
\end{equation}
where the tilde implies the unphysical case $\JD\neq 0$.
Using the physical case $\JD=0$ given in (\ref{gC})-(\ref{gA}),
the components of (\ref{Gotil}) can be written as
\begin{eqnarray}
  \gCtil[\JD,\JC]
  &\equiv& \lra{ 1-\gC[\JC]\frac{\JD}{\ri\hbar} }^{-1} \gC[\JC]  
,
\label{gCtil}
\\
  \gRtil[\JD,\JC] 
  &\equiv& \lra{ 1+\gCtil[\JC,\JD]\frac{\JD}{\ri\hbar} } \gR  
,
\\
  \gAtil[\JD,\JC] 
  &\equiv& \gA \lra{ 1+\frac{\JD}{\ri\hbar}\gCtil[\JC,\JD] } 
,
\\
  \gDtil[\JD,\JC] 
  &\equiv& \gA 
           \lra{ \frac{\JD}{\ri\hbar}
                -\frac{\JD}{\ri\hbar}\gCtil[\JC,\JD]
                 \frac{\JD}{\ri\hbar} 
               } 
           \gR
,
\label{gDtil}
\end{eqnarray}
with a short-hand notation.
Of course $\Gotil$ is reduced to $\Go$ in (\ref{Go}) by setting $\JD=0$.
Note that retarded or advanced nature of $\gR$ or $\gA$,
respectively,
is recovered only in the physical case $\JD=0$.

Then the generating functional $W$ can be expressed as
\begin{equation}
  \frac{\ri}{\hbar}W[J,\ID] = \Tr \ln \Gotil[J] +\kappa[J,\ID]
,
\label{W2}
\end{equation}
where $J$ expresses the set of $\JD$ and $\JC$,
and $\kappa$ is the sum of all the connected diagrams 
constructed by the propagator $-\Gotil[J]$, 
the vertex $V$ and the external leg $\ID$.
For simplicity,
we suppress the argument $\ID$ in this subsection.

Next we evaluate (\ref{W2}) at $J=\Jo[z]+\DelJ[z]$ (cf.~(\ref{J2})),
and substitute it 
into the definition (\ref{Gamma0}) of the effective action $\mG$.
Expanding $\mG$ around $J=\Jo[z]$ in terms of $\DelJ[z]$,
the terms linear in $\DelJ$ are canceled,
and we obtain
\begin{eqnarray}
  \frac{\ri}{\hbar}\mG[z] 
  &=& \Tr \ln \Gotil[\Jo+\DelJ] +\kappa[J[z]]
     -\frac{\ri}{\hbar}
      \lrb{\lra{ \JoD+\DelJD}\zC +\lra{\JoC+\DelJC}\zD }
\\
  &=& \frac{\ri}{\hbar}\mGo[z] 
     -\frac{1}{2\hbar^{2}}
      \begin{array}{c}
        \lra{ \DelJD   \DelJC } \\
        {}                      \\
      \end{array}
      \Delta_{2}
      \lra{\begin{array}{c}
        \DelJD \\
        \DelJC \\
      \end{array}
      }
     +\kappa'[z]
,
\label{Gamma2}
\end{eqnarray}
where $\mGo$, $\Delta_{2}$ and $\kappa'$ 
are defined respectively by
\begin{eqnarray}
  \frac{\ri}{\hbar}\mGo[z] 
  &\equiv& \Tr \ln \Gotil[\Jo] 
          -\frac{\ri}{\hbar}\lrb{ \JoD\zC +\JoC\zD }
,
\\
  \Delta_{2}(t,s)
  &\equiv&
     -\lra{\begin{array}{cc}
             \gCtil(t,s)\gCtil(s,t)    &  \ri\gRtil(t,s)\gAtil(s,t) \\
             \ri\gAtil(t,s)\gRtil(s,t) & -\gDtil(t,s)\gDtil(s,t)    \\
           \end{array}
           }
,
\label{Del2def}
\\
  \kappa'[z]
  &\equiv& \kappa[J[z]] 
          +\Tr\sum_{k\geq 3}\frac{1}{k} 
           \lrb{ \frac{1}{\ri\hbar}\Gotil[\Jo] 
                 \lra{\begin{array}{cc}
                       \DelJD & 0         \\
                       0      & \ri\DelJC \\
                      \end{array}
                     }
               }^{k}
.
\end{eqnarray}
Eq.~(\ref{Gamma2}) corresponds to Eq.~(3.21) in \Ref{Yokojima},
and then,
as it is proved in \Ref{Yokojima},
the effective action can be expressed as
\begin{equation}
  \frac{\ri}{\hbar}\mG[z]
   =  \cR_{2}\lra{\frac{\ri}{\hbar}W[\Jo[z]]}
     -\frac{\ri}{\hbar}\sum_{\vk,\vq}\intIF\rd t
      \lra{ \JokqD[z]\zkqC+\JokqC[z]\zkqD }
.
\label{Gamma3}
\end{equation}
Here,
$\cR_{2}$ is a diagrammatic operation defined by the following process.
\begin{enumerate}
\item
\label{2PRcutJoint}
The first process of $\cR_{2}$ can be expressed schematically as
\begin{equation}
 \raisebox{-3.5mm}{\epsfile{width=22mm,file=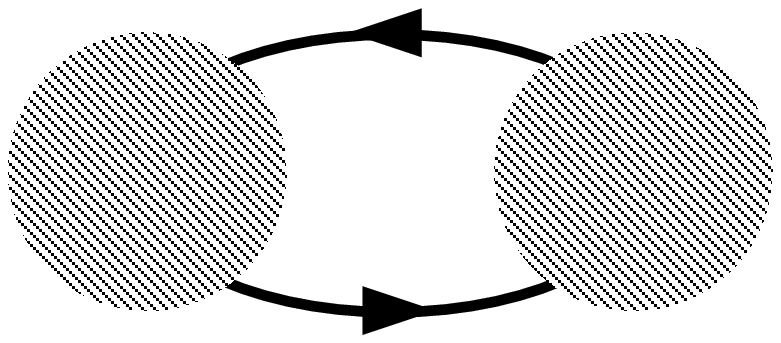}} 
  \Longrightarrow
  \raisebox{4mm}{$
  \lra{\raisebox{-4mm}{\epsfile{width=29.7mm,file=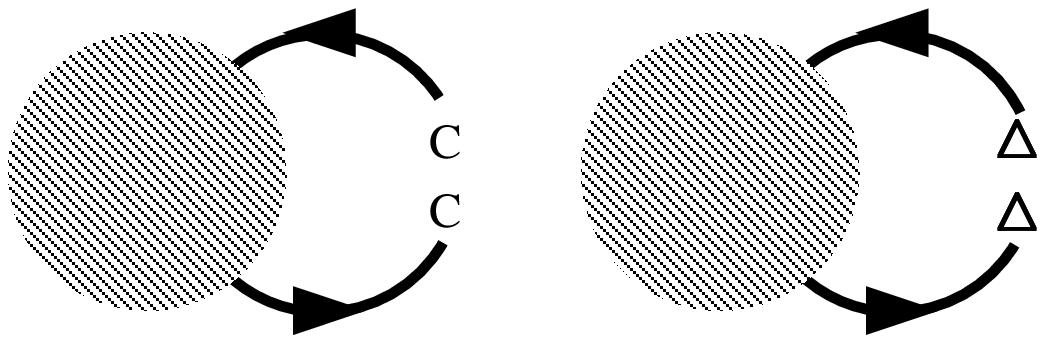}}}
  $}
  \Delta_{2}^{-1}
  \lra{\raisebox{-10mm}{\epsfile{width=13.1mm,file=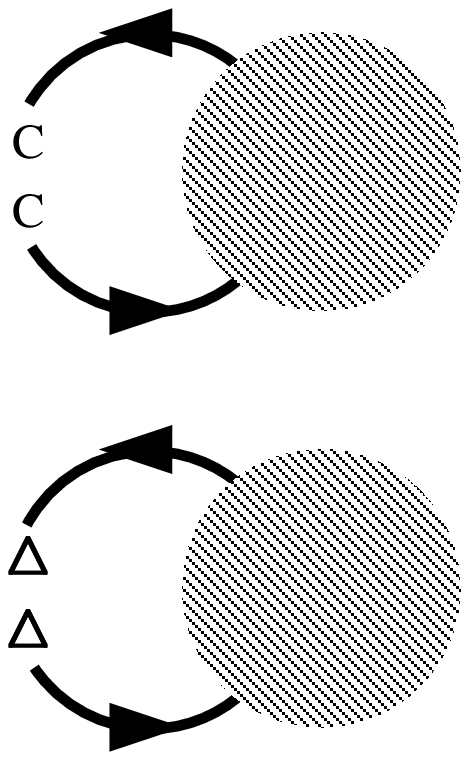}}} 
,
\label{cutpatch}
\end{equation}
to which we refer as the `cut-and-patch' operation:
If there is a 2PR part in the diagram,
separate the graph into two pieces 
by cutting the corresponding pair of the propagators.
In each of the separated diagrams,
make the resultant two external lines to contract
$\psiC^{\ast}(t)\psiC(t)$ or $\ri\psiD^{\ast}(t)\psiD(t)$,
which we call the  $\zC$- or $\zD$-leg,
respectively.
Then reconnect the two diagrams 
by contracting their $z$-legs with $\Delta_{2}^{-1}$.
\item
The second step is to carry out the procedure \ref{2PRcutJoint} 
in all possible ways,
and sum up all the resulting diagrams including the original one.
\end{enumerate}
For example,
\begin{equation}
  \cR_{2}\lra{ \raisebox{-5mm}
                        {\epsfile{width=10.65mm,file=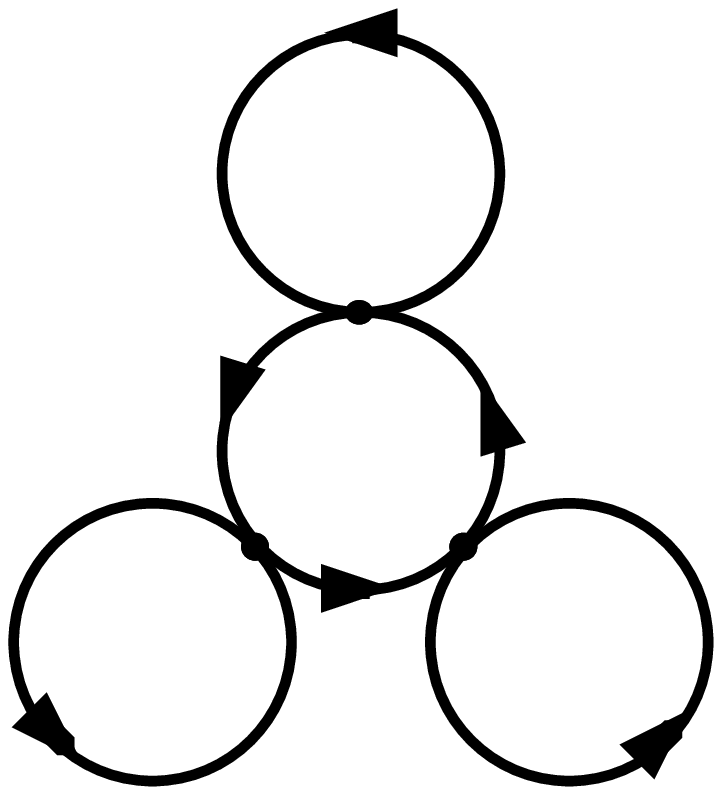}}
              }
   =  \raisebox{-5mm}{\epsfile{width=10.65mm,file=R2-0.eps}} 
     +\raisebox{-6mm}{\epsfile{width=14.75mm,file=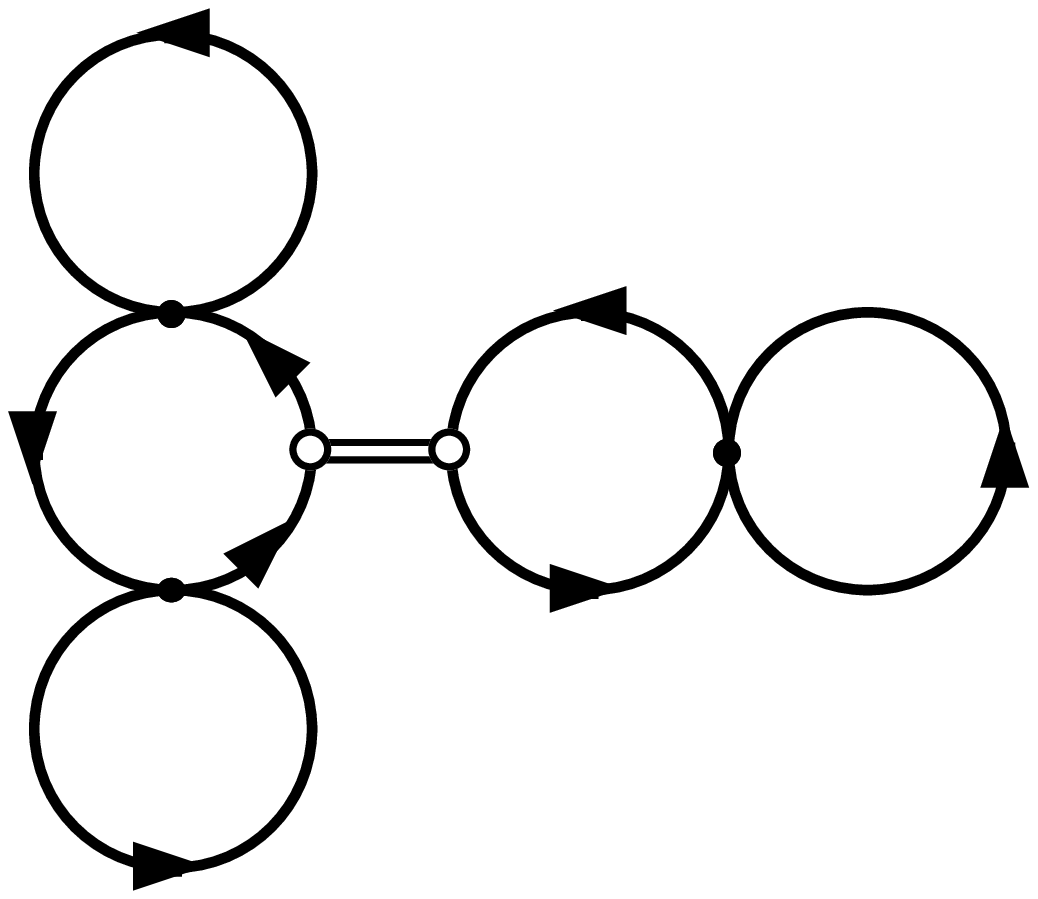}}
     +\raisebox{-7mm}{\epsfile{width=12.75mm,file=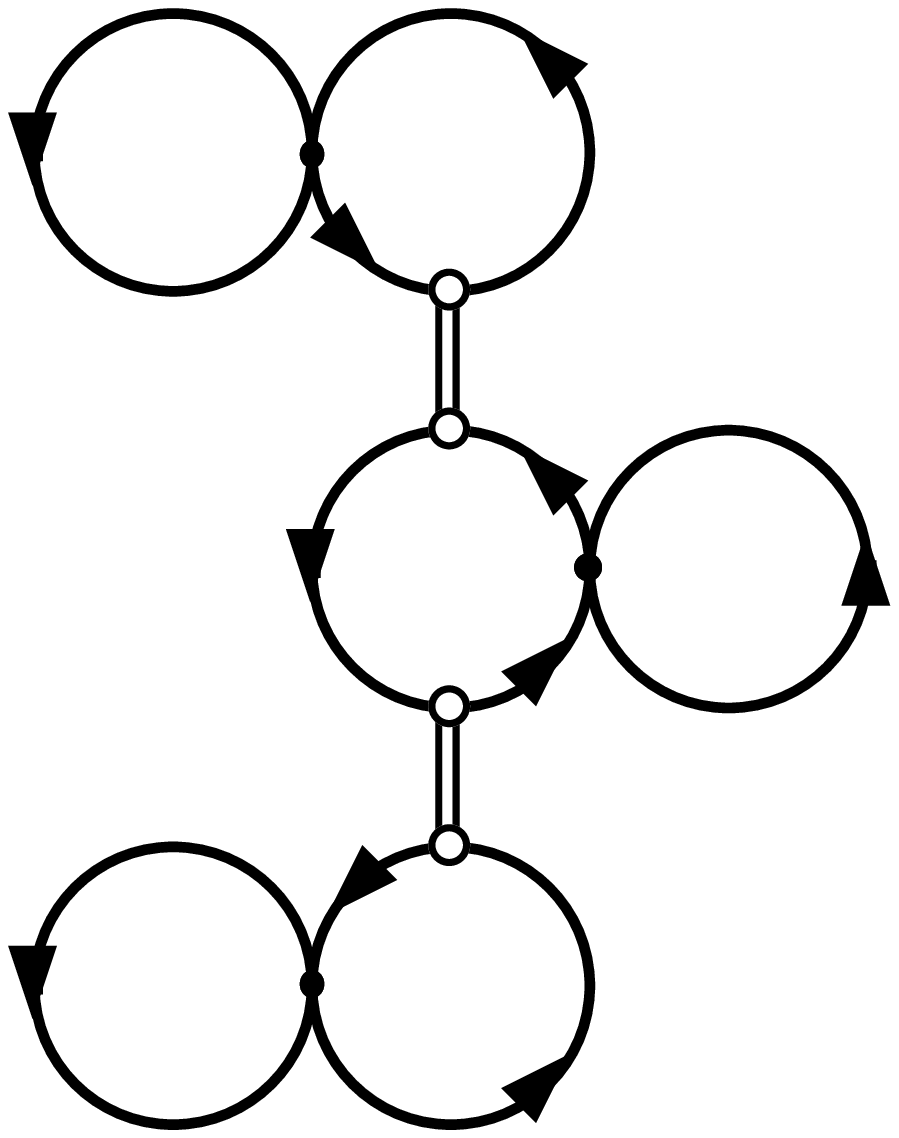}}
     +\raisebox{-7mm}{\epsfile{width=18.75mm,file=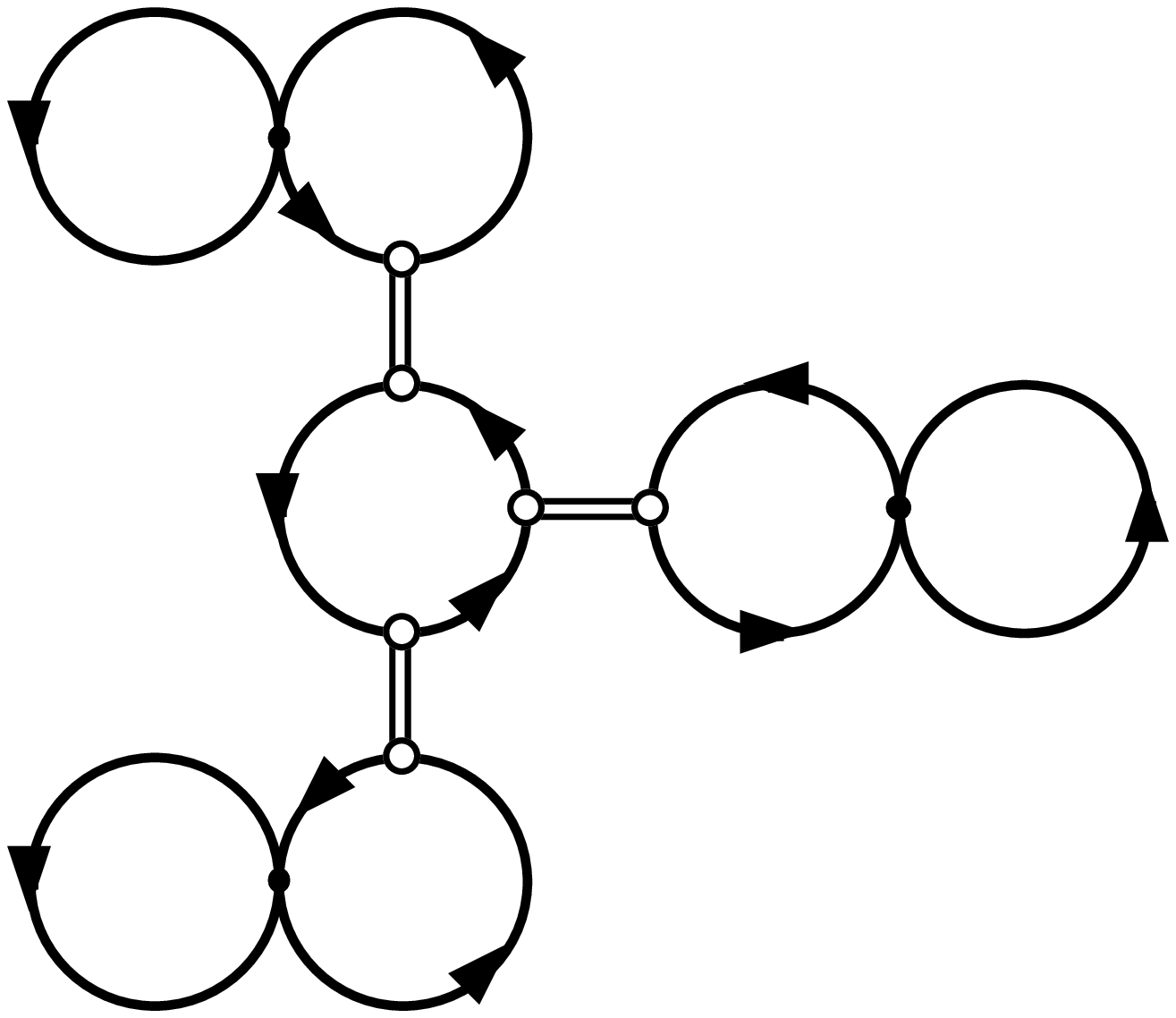}}
,
\end{equation}
where $\rDel_{2}^{-1}$ is expressed by doubled lines.
It should be emphasized 
that $W$ is evaluated at $J=\Jo[z]$ in (\ref{Gamma3}):
In the diagrammatic expression of $W$,
the propagators are $-\Go[\Jo[z]]$ 
which acquire the form of the GKB ansatz when we set $\zD=0$.

As it was discussed in \Ref{Yokojima},
the operation $\cR_{2}$ cancels some part of the 2PR diagrams,
and in this sense,
$\mG$ has a modified 2PI property.
It is reduced to the usual 2PI 
when we discuss an effective action 
of non-local operator $\psid(t)\psi(s)$.
In the next subsection,
we will clarify what is the contents of this modified 2PI property.

\subsection{Diagrammatic Rule for $Q[\zC]$}
\label{Q-rule}

Recovering the argument $\ID$ in (\ref{Gamma3}),
the expectation value $Q$ as a functional of the WDF 
is obtained from (\ref{Q2}) as
\begin{equation}
  Q[t;\zC]
   =  \rbar{ \cR_{2}\lra{ \frac{\delta W[\Jo[z],\ID]}{\delta\ID(t)} } }
           { \zD=\ID=0 }
.
\label{Q3}
\end{equation}
Note that $\Jo[z]$ in (\ref{Gamma3}) does not depend on $\ID$.
Diagrammatically,
inside the operation $\cR_{2}$ is a sum of all the connected diagrams
with one external point expressing $Q(\psiC)$.
For definiteness,
we consider the case 
$\hat{Q} = \psid_{\vq}\psid_{\vqq}\psio_{\vkk}\psio_{\vk}$
as an example.
In the following,
since we have set $\zD=\ID=0$ in (\ref{Q3}),
the form of the Green function $\Gotil[\Jo[z]]$ 
is reduced to that of the GKB ansatz;
Eq.(\ref{Go}) in which $\zo[\JC]$ is replaced by $z$.

\subsubsection{Time-ordered configuration and causality}
\label{Causality}

Before considering the operation $\cR_{2}$,
we define terminologies 
`time-ordered configuration' and `causality'.

\begin{figure}[htb]
\begin{center}
\raisebox{6mm}{ \epsfile{width=11.5mm,file=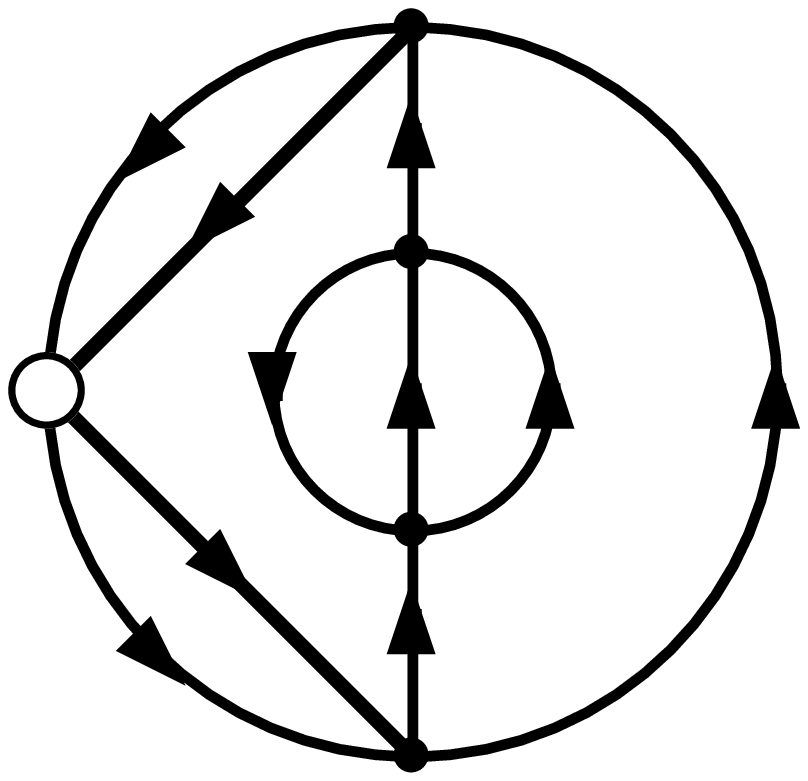} }
\hspace{1em}
\epsfile{width=16.8mm,file=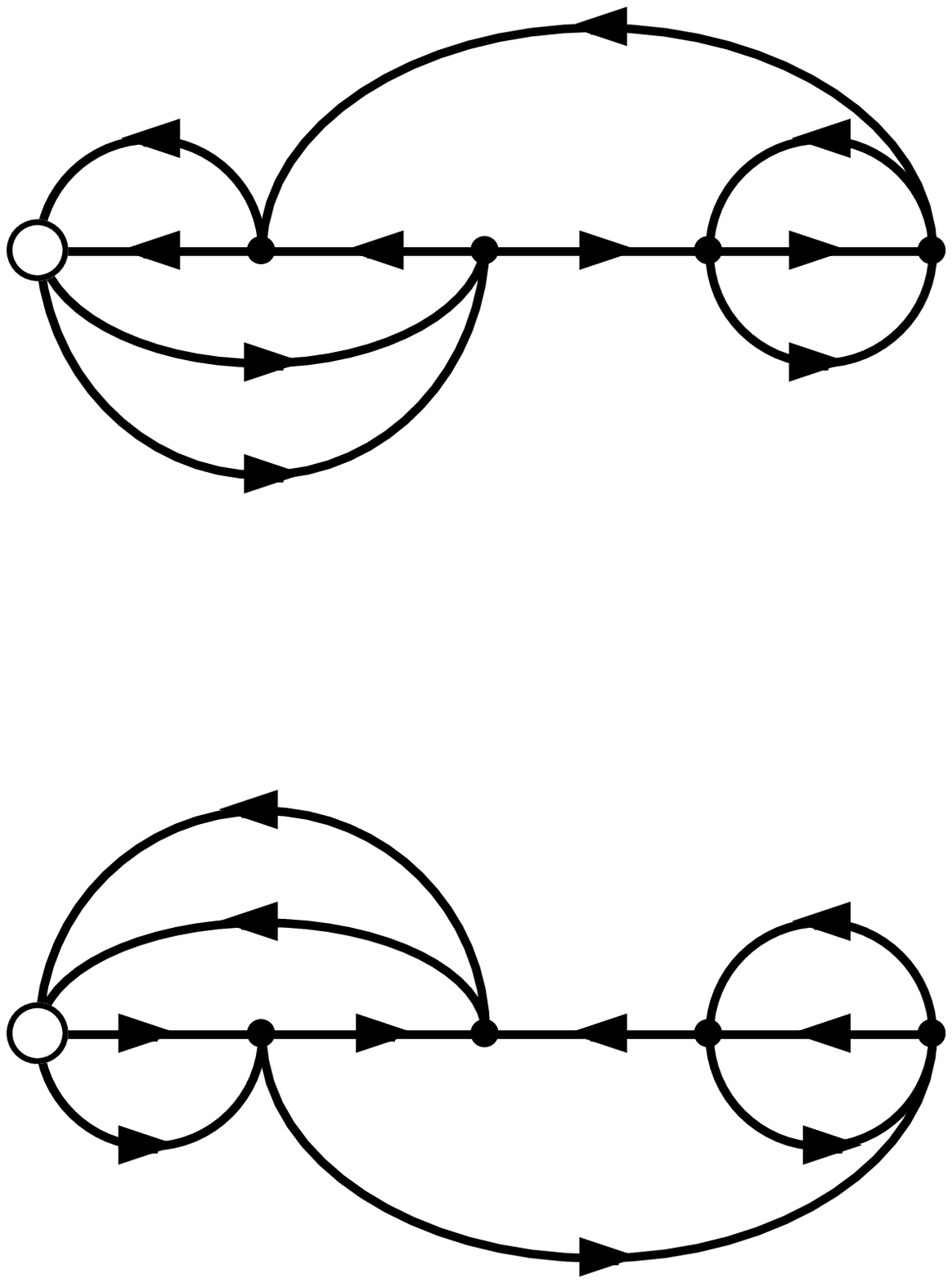}
\epsfile{width=16.8mm,file=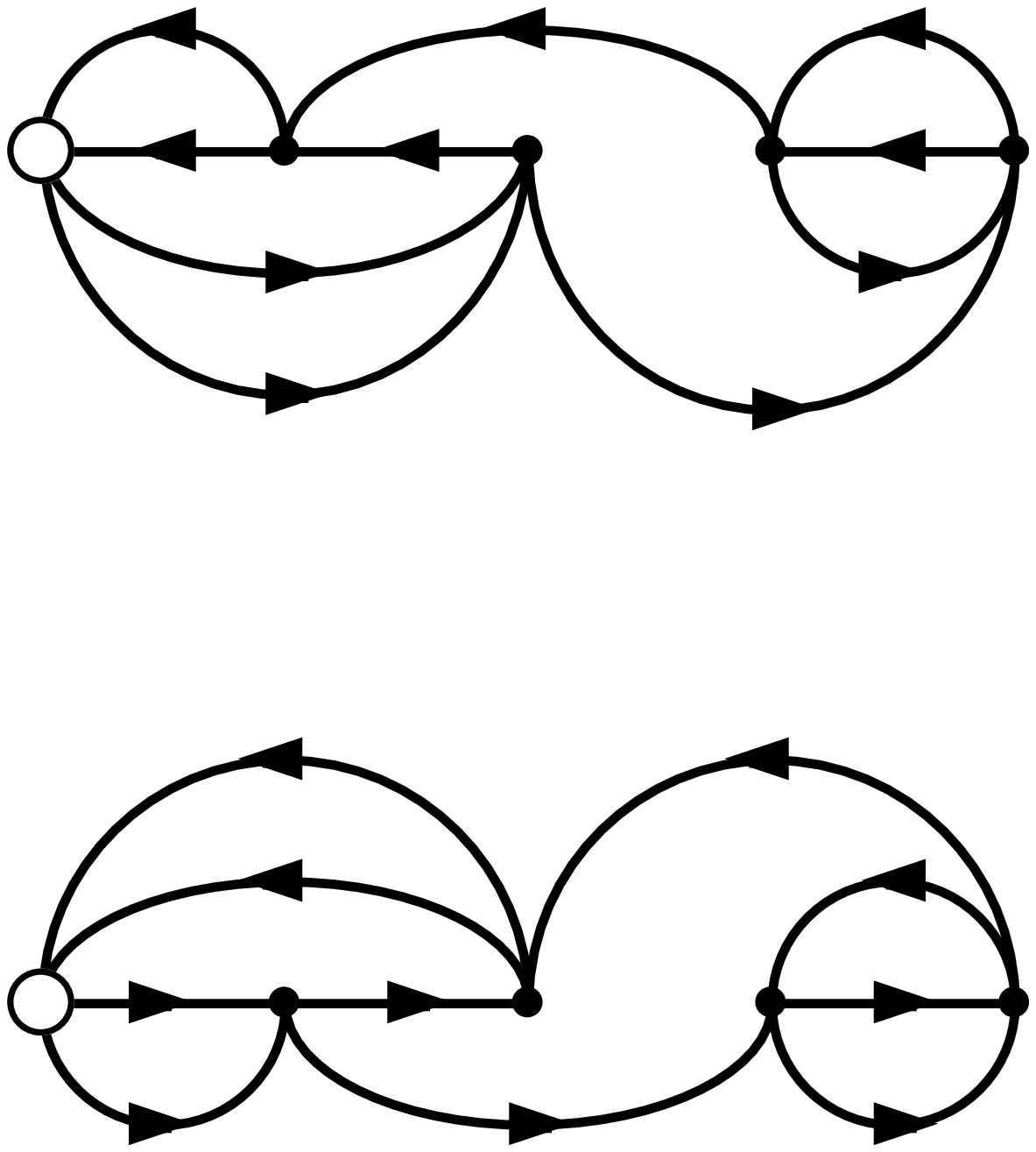}
\hspace{1em}
\raisebox{-1.6mm}{
\epsfile{width=16.8mm,file=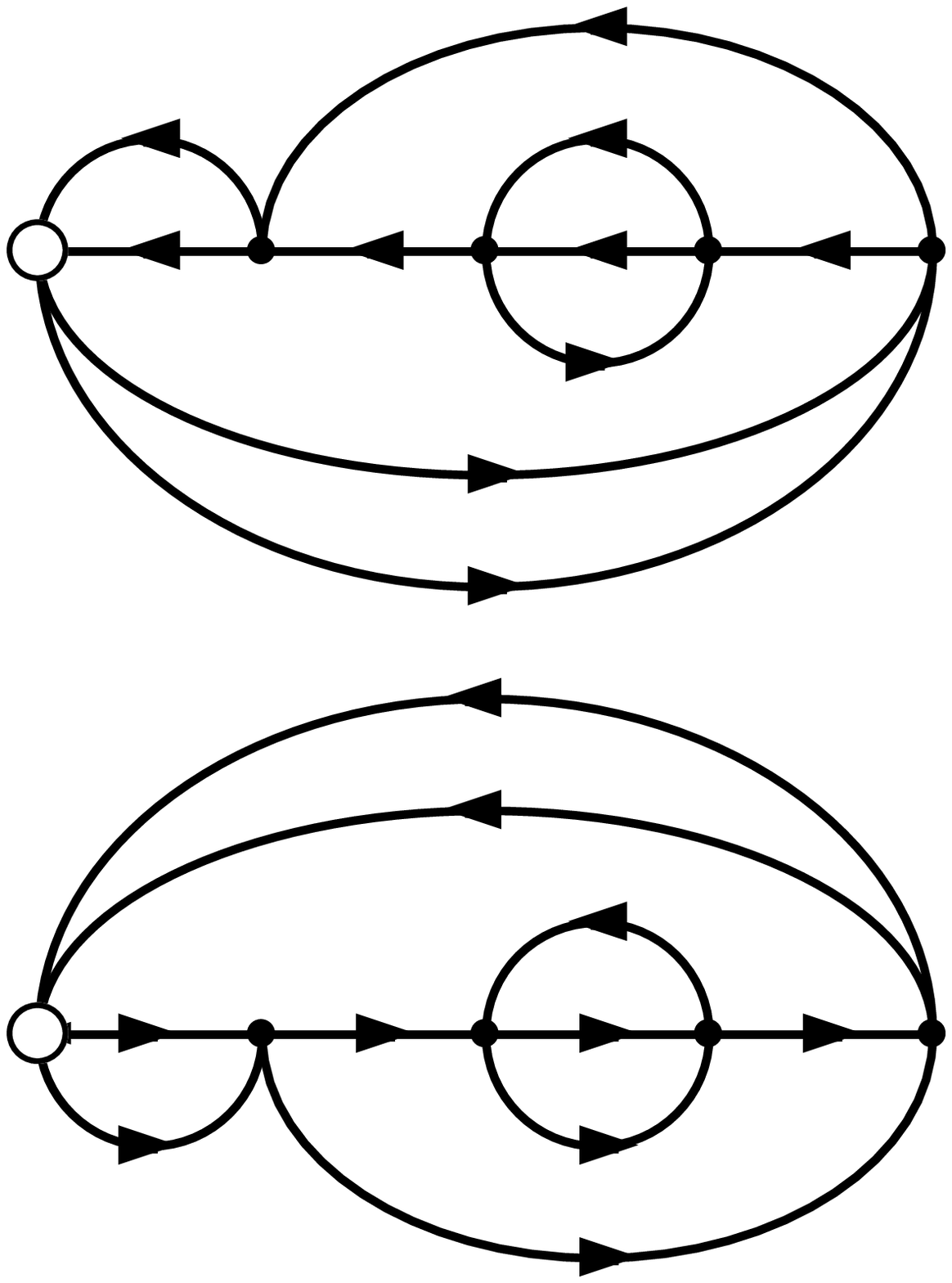}
\epsfile{width=16.8mm,file=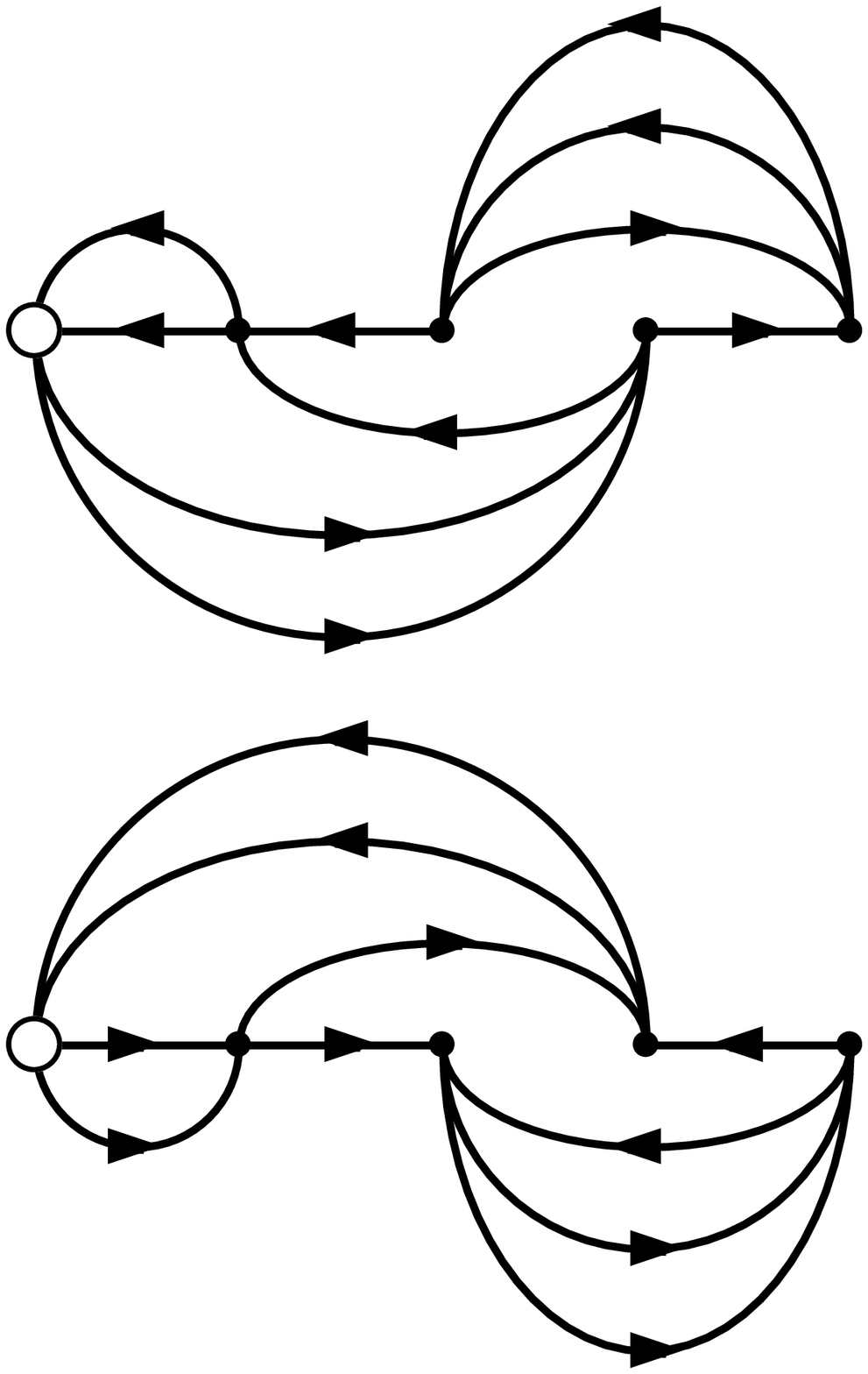}
}
\\
\makebox[11.5mm][l]{{\footnotesize (a)}}
\hspace{1em}
\makebox[34mm][l]{{\footnotesize (b)}}
\hspace{1em}
\makebox[34mm][l]{{\footnotesize (c)}}
\end{center}
\caption{
Examples of time-ordered configurations.
The open circle expresses the external point $Q$.
The diagram (a) can be arranged 
into eight configurations shown in (b) and (c).
The difference between the groups (b) and (c) will be clarified later.
}
\label{L4ex}
\end{figure}

In the non-equilibrium Green function technique,
because the propagator depends explicitly on time,
the evaluation of a diagram may be carried out 
as a function of time as follows.
For all possible ways of time ordering of the vertices,
the diagram is arranged in such a way that the vertices are put 
on the time axis from right to left.
Then assigning the factors of propagators and vertices,
each time ordering gives different contribution.
In the following,
we refer to the diagram with a fixed time ordering of the vertices
as the `time-ordered configuration' or simply the configuration.
For example,
the diagram in \figref{L4ex}~(a) can be arranged 
as the eight configurations shown in \figref{L4ex}~(b) and (c).
Other possible configurations
can be eliminated by the following mechanism.

\begin{figure}[htb]
\begin{center}
\epsfile{width=3cm,file=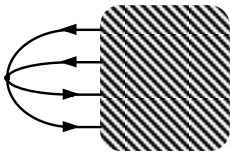}
\end{center}
\caption{
Vanishing configuration due to the causality.
(For definiteness,
 the four point interaction in (\protect\ref{Hint}) is considered.
)
}
\label{causal}
\end{figure}

The vertices in the diagram expresses 
$H_{\rm int}(\psi_{1})-H_{\rm int}(\psi_{2})$ 
rewritten by $\psiD$ and $\psiC$,
which is odd in $\psiD$ for generic $H_{\rm int}$,
and contains at least one $\psiD$ or $\psi_{\rDel}^{\ast}$.
Then,
we can conclude that 
the time-ordered configuration like \figref{causal},
where a vertex is on the latest time,
vanishes:
Assuming the vertex of \figref{causal} is on time $t$,
$\psiD(t)$ or $\psi_{\rDel}^{\ast}(t)$ therein
must be contracted by $\gA(t,s)$ or $\gR(s,t)$ ($t>s$),
respectively,
(Recall that we are working in the physical case $\JD=0$.)
and their advanced or retarded character leads to the vanishing.
This implies that,
when we calculate an expectation value 
of a physical quantity at time $t$,
the interaction at time later than $t$ does not contribute
since the configuration like \figref{causal} can not be avoided.
In other word,
the time-ordered configuration of the diagram for $Q$ 
must have the external point $Q$ 
on the latest time within the diagram.
In this article,
we call such a fact the `causality'.

\subsubsection{Meaning of the operation $\cR_{2}$}

Now we consider the meaning of the operation $\cR_{2}$ in (\ref{Q3}).
For this sake,
we first examine the cut-and-patch operation.
Since we are considering the physical case $\zD=0$,
the $\rDel\rDel$-component of $\Gotil$ 
as well as the $\rDel\rDel$-component of $\rDel_{2}$ vanish.
Then,
because $\rDel_{2}^{-1}$ can be written as
\begin{eqnarray}
  \rDel_{2}^{-1}
  &=&-\lra{ \begin{array}{cc}
              \gC\gC    & \ri\gR\gA \\
              \ri\gA\gR & 0         \\
            \end{array}
           }^{-1}
\nonumber\\
  &=& \lra{ \begin{array}{cc}
              0                    & 
              \ri\{\gA\gR \}^{-1}
            \\
              \ri\{\gR\gA \}^{-1} & 
             -\{ \gR\gA \}^{-1} \gC\gC \{ \gA\gR \}^{-1} 
            \\
            \end{array}
           } 
,
\label{Del2inv}
\end{eqnarray}
the $\rC\rC$-component of $\rDel_{2}^{-1}$ disappears.
Thus,
in (\ref{cutpatch}),
the connection of two $\zC$-legs is absent.

\begin{figure}[htb]
\begin{center}
\epsfile{width=5cm,file=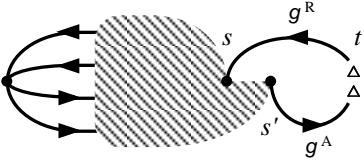}
\end{center}
\caption{
The time-ordered configuration 
of sub-diagram with a $\zD$-leg but without the external point.
(for convenience, the time $s$ is chosen to be later than $s'$.)
}
\label{zDleg2}
\end{figure}

Moreover,
in (\ref{Q3}),
the connection of two $\zD$-legs is forbidden 
by the following reason.
Since there is only one external point $Q$
in each diagram for (\ref{Q3}),
the external point belongs to only one of the two sub-diagrams
connected by $\rDel_{2}^{-1}$.
Then,
if both of the two sub-diagrams are connected with their $\zD$-legs,
the one which does not contain the external point $Q$ 
must vanish due to the causality:
The $\zD$-leg at time $t$ is produced 
by a pair of Green functions $\gR(s,t)$ and $\gA(t,s')$,
which implies they must be connected to vertices
at time $s$ and $s'$ later than $t$.
So,
if the sub-diagram does not have an external point,
any time-ordered configuration of the sub-diagram
must have a vertex which possesses at the latest time 
as shown in \figref{zDleg2},
and this gives a vanishing contribution due to the causality
discussed in the previous subsection.

\begin{figure}[htb]
\begin{center}
\raisebox{-4mm}{ \epsfile{width=22.3mm,file=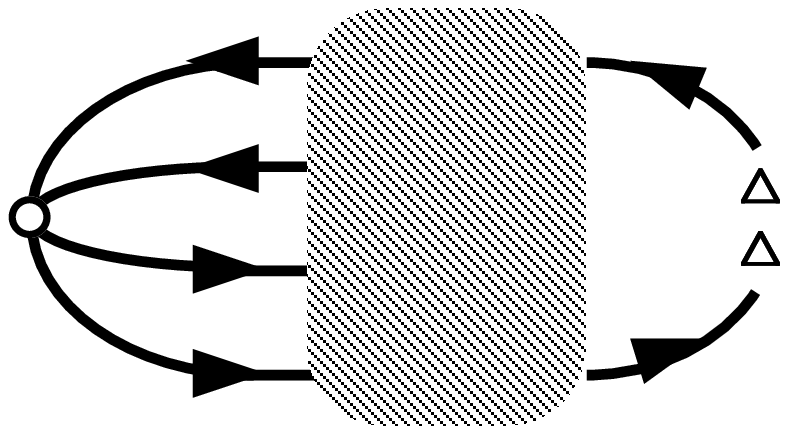} }
\mbox{$\ri\lra{\gR\gA}^{-1}$}
\raisebox{-4mm}{ \epsfile{width=14.1mm,file=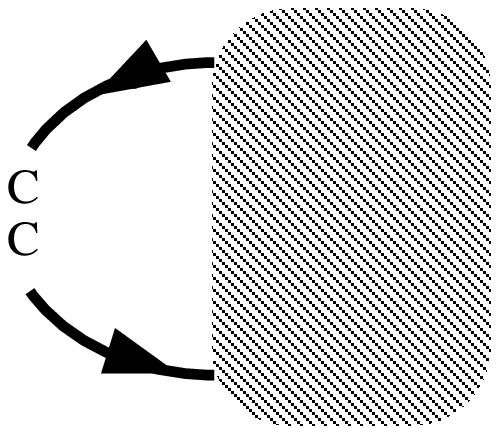} }
\end{center}
\caption{
The generic structure of the diagram for $Q[z]$.
}
\label{Q4}
\end{figure}

Thus it is enough to consider the connection of $\zD$-leg and $\zC$-leg,
where the external point belongs to the sub-diagram with $\zD$-leg.
In contrast to $\zD$-leg,
the $\zC$-leg must be possessed 
on the latest time within the sub-diagram:
Otherwise,
some vertex must be possessed on the latest time 
because the sub-diagram with $\zC$-leg does not have the external point,
and the configuration in \figref{causal} cannot be avoided.
As the result,
the time-ordered configurations for $Q$ has a generic form 
shown in \figref{Q4}:
In the sub-diagram with $\zD$-leg,
the external point $Q$ is on the latest time,
and the $\zD$-leg is connected to the vertices on the later time 
(need not be on the  earliest time of the sub-diagram).
On the other hand,
the sub-diagram with $\zC$-leg has the leg on the latest time.

Let us see the joint of $z_{\vkk\vk,\rDel}$-leg at $t$ 
and $z_{\vq\vqq,\rC}$-leg at $s$
by $\ri\{\gR\gA \}^{-1}_{\vk\vkk,\vq\vqq}(t,s)$.
The $z_{\vkk\vk,\rDel}$-leg at time $t$ 
is produced by a pair of propagators which can be written as
\begin{eqnarray}
  \ri \gR_{\wk,\vk}(t',t) \gA_{\vkk,\wkk}(t,t'')
  &=&  \ri \theta(t'-t'')\re^{-\ri\omega_{\wk}(t'-t'')}
       \gR_{\wk,\vk}(t'',t) \gA_{\vkk,\wkk}(t,t'')
\nonumber\\
  & & +\ri \theta(t''-t')\re^{\ri\omega_{\wkk}(t''-t')}
       \gR_{\wk,\vk}(t'',t) \gA_{\vkk,\wkk}(t,t'')
.
\label{gRgA}
\end{eqnarray}
On the other hand,
the $z_{\vq\vqq,\rC}$-leg at time $s$ is produced 
by a pair of propagators,
one of which is $-\gR_{\vq,\wq}(s,s')$ or $-\gC_{\vq,\wq}(s,s')$ 
and the other is $-\gA_{\wqq,\vqq}(s'',s)$ or $-\gC_{\wqq,\vqq}(s'',s)$.
As discussed above,
$\zC$-leg is non-zero only when $s > s',s''$,
and the following relations hold in this case;
\begin{eqnarray}
  \theta(t'-s)\re^{-\ri\omega_{\vq}(t'-s)} 
  g^{{\rm R}/\rC}_{\vq,\wq}(s,s')
  &=& g^{{\rm R}/\rC}_{\vq,\wq}(t',s')
,
\label{gRC}
\nonumber\\
  \theta(t''-s)\re^{\ri\omega_{\vqq}(t''-s)} 
  g^{{\rm A}/\rC}_{\wqq,\vqq}(s'',s)
  &=& g^{{\rm A}/\rC}_{\wqq,\vqq}(s'',t'')
.
\label{gAC}
\end{eqnarray}
With the aid of Eqs.~(\ref{gRgA})-(\ref{gAC}),
we have
\begin{equation}
  \sum_{\vk,\vkk,\vq,\vqq} \int \rd t \rd s\,
  \ri \gR_{\wk,\vk}(t',t) \gA_{\vkk,\wkk}(t,t'')
      \ri \{-\gR\gA \}^{-1}_{\vk\vkk,\vq\vqq}(t,s)\,
      g^{{\rm R}/\rC}_{\vq,\wq}(s,s')\,
      g^{{\rm A}/\rC}_{\wqq,\vqq}(s'',s)
   = -g^{{\rm R}/\rC}_{\wk,\wq}(t',s')
      g^{{\rm A}/\rC}_{\wqq,\wkk}(s'',t'')
,
\end{equation}
where $t',t''>s',s''$ holds.
This implies that
the joint of $\zD$- and $\zC$-legs can simply be expressed as
\begin{equation}
\raisebox{-4mm}{\epsfile{width=13mm,file=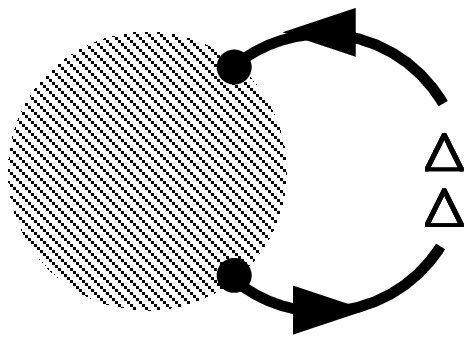}}
\:\ri \{\gR\gA \}^{-1}\:
\raisebox{-4mm}{\epsfile{width=13mm,file=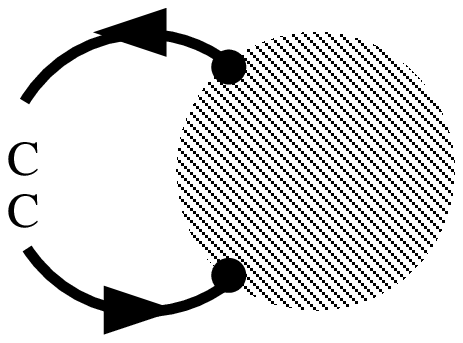}}
=
-\,
\raisebox{-4mm}{\epsfile{width=24mm,file=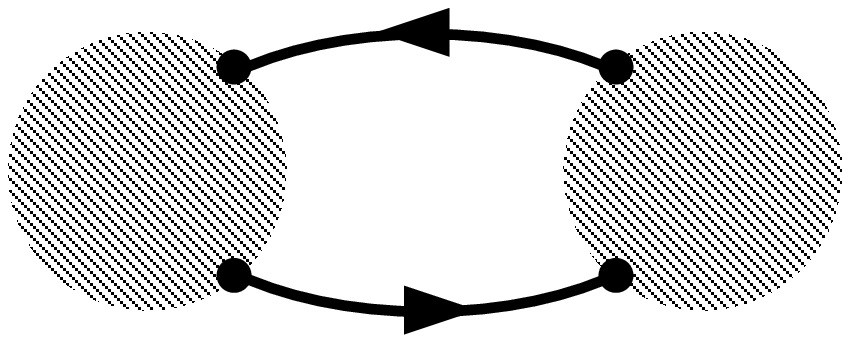}}
,
\label{zD-zC}
\end{equation}
Note that,
on the rhs of (\ref{zD-zC}),
the time order of the vertices is restricted
unlike usual diagrams:
The vertices originally connected to the $\zD$-leg
are on later time than those originally connected to $\zC$-leg.
This implies that
the diagram on the rhs of (\ref{zD-zC}) contains 
only the configurations which can be separated into two parts
by cutting the pair of propagators at the same instant.
In this sense,
we call such a time-ordered configuration 
`instantaneous-2PR configuration'.
For instance,
the configurations shown in \figref{L4ex}~(b) are instantaneous-2PR,
and those in (c) are instantaneous-2PI.
Summarizing,
the cut-and-patch operation (\ref{cutpatch})
extracts an instantaneous-2PR configuration 
from the original diagram with the opposite signature.

The second process of $\cR_{2}$ is 
to operate cut-and-patch in all possible ways 
and to sum up the resultant diagrams.
This process ensures 
that the instantaneous-2PR configuration is precisely canceled out
after $\cR_{2}$ is carried out.
As it was shown above,
the cut-and-patch process
just restricts the time ordering of the vertices,
and except the signature,
the contribution produced by cut-and-patch 
is included in the original diagram.
Then we should count
how many times the same contribution appears throughout $\cR_{2}$.
Considering a configuration which 
is instantaneous-2PR with respect to $N$ pairs of propagators,
such a configuration appears in a diagram 
where $k$ of the corresponding $N$ pairs of the propagators 
are cut-and-patched.
There are ${}_{N}{\rm C}_{k}$ ways of choosing $k$ pairs,
and the signature $(-1)^{k}$ is assigned.
Thus through the total process of $\cR_{2}$,
the instantaneous-2PR configuration appears 
$\sum_{k}(-1)^{k}{}_{N}{\rm C}_{k}=0$ times.

As the result,
\begin{quote}
the operation $\cR_{2}$ on a diagram implies 
that we can eliminate the instantaneous-2PR configurations
which will be produced from the original diagram.
\end{quote}
For example,
considering (\ref{Q3}) in the fourth order of the perturbation,
when we evaluate the diagram shown in \figref{L4ex}~(a),
we only need to calculate the contributions 
of the instantaneous-2PI configurations shown in \figref{L4ex}~(b),
and can eliminate the instantaneous-2PR ones in (b).
More simple example can be seen in \Ref{Koide3}
(explicit calculations are shown in \Ref{thesis}),
where the four-point function 
is calculated up to the first order of the perturbation.
There appear tadpole diagrams,
but their contributions are canceled 
when the four-point function is expressed in terms of the WDF.
From the view point of our rule,
the contribution from the tadpole diagrams in \Ref{Koide3} 
can be eliminated by the $\cR_{2}$ operation in (\ref{Q3}) 
because all of the time-ordered configurations produced 
from those diagrams are instantaneous-2PR.

\subsection{Quantum kinetic equation}
\label{QKE-rule}

Finally,
we summarize the rule for deriving the QKE.
The physical source $\JC$ as a functional of $\zC$ is obtained 
by setting $\zD=0$ in the first equation of (\ref{Jderiv}) 
since this condition is equivalent to $\JD=0$.
With the use of (\ref{Gamma3}),
it can be expressed as
\begin{eqnarray}
  \JC[t;\zC]
  &=& \JoC[t;\zC]
   -  \int \rd s 
      \rbar{ \frac{\delta \JoD(s)}{\delta \zD(t)} }{ \zD=0 }
      \lrb{ \cR_{2}\lra{ \frac{\delta W}{\delta \JD(s)}
                        }_{ J=\Jo[\zD=0] }
           -\zC(s)
          }
,
\label{J3}
\end{eqnarray}
and the QKE for the WDF $\zC$ is obtained by setting $\JC[t;\zC]=0$.

To obtain the explicit expression 
of $\delta\JD/\delta\zD$ in (\ref{J3}),
we differentiate the identity $z=\zo[\Jo[z]]$ with respect to $z$,
and obtain
\begin{equation}
  \lra{ \begin{array}{cc}
          \frac{\delta \JoD}{\delta \zC} &
          \frac{\delta \JoD}{\delta \zD}
        \\
          \frac{\delta \JoC}{\delta \zC} &
          \frac{\delta \JoC}{\delta \zD}
        \end{array}
       }_{\zD=0}
  = \lra{ \begin{array}{cc}
            \frac{\delta \zC^{(0)}}{\delta \JD} &
            \frac{\delta \zC^{(0)}}{\delta \JC}
          \\
            \frac{\delta \zD^{(0)}}{\delta \JD} &
            \frac{\delta \zD^{(0)}}{\delta \JC}
          \end{array}
         }^{-1}_{ J=\Jo[\zD=0] }
.
\label{dJ/dz}
\end{equation}
Because $\zC^{(0)}[t;\JD,\JC]$ and $\zD^{(0)}[t;\JD,\JC]$ are given 
by $-\gCtil(t,t)$ and $-\ri\gDtil(t,t)$,
respectively,
their derivatives can be calculated 
using the definitions (\ref{gCtil}) and (\ref{gDtil}).
Then we can see that the rhs of (\ref{dJ/dz})  is nothing but
$\rDel_{2}^{-1}$ (multiplied by $\ri\hbar$) given in (\ref{Del2inv}),
and $\delta\JoD/\delta\zD$ is reduced to
\begin{eqnarray}
  \rbar{ \frac{\delta \Jo_{\vqq\vkk,\rDel}(s)}{\delta \zkqD(t)} }
       { \zD=0 }
  &=&-\hbar\lrb{\gA\gR}^{-1}_{\vk\vq,\vkk\vqq}(t,s)
\nonumber\\
  &=&-\lrb{ \hbar\del_{t}+\ri\lra{\epk-\epq} }
      \delta_{\vk,\vkk}\delta_{\vq,\vqq}\delta(t-s)
.
\end{eqnarray}

Thus,
in the rhs of (\ref{J3}),
the last term in the braces compensates 
for the first term (cf.~(\ref{J0})),
and the QKE can simply be written as
\begin{equation}
   \lrb{ \hbar\del_{t}+\ri\lra{\epk-\epq} }
   \lrb{ \cR_{2}\lra{ \zkqC[t;\JC=\JoC[\zC]] } }
   = 0
.
\label{QKE}
\end{equation}
The QKE can be derived 
by calculating the instantaneous-2PI configurations 
of the diagrams for $\zkqC$,
and by operating $\lrb{ \hbar\del_{t}+\ri\lra{\epk-\epq} }$.
This rule can be confirmed 
by the example in \Ref{Koide3} (details are in \Ref{thesis}).
There,
it is explicitly shown that the tadpole diagrams for $\zC$ are 
canceled through the process of inversion.
According to our rule,
the tadpole diagrams in \Ref{Koide3} must vanish
because they necessarily lead to instantaneous-2PR configurations.

\section{Discussions}

We have presented a systematic method to 
to calculate an expectation value $Q(t)$ 
of some physical quantity $\hat{Q}$ as a functional of the WDF $z$.
Using the propagator $-\Go[\Jo[z]]$,
which has a form of the GKB ansatz,
the precise expression of $Q[z]$ is obtained by
eliminating the instantaneous-2PR configurations from the calculations.
This is due to a restriction which must be taken into account
in the course of the perturbative calculation:
the integration over the microscopic field variable 
must be carried out in a way so that the value of the WDF is fixed.
Together with the QKE,
which can also be expressed in a compact form 
by the use of instantaneous-2PI property (cf.~(\ref{QKE}),
this method provides us 
with a complete framework for the quantum kinetic theory.

As pointed out in \secref{problem},
the method presented here can straight forwardly be used 
in the GKB formalism.
What we have used for the propagator 
is a GKB ansatz with the free-particle approximation 
of the spectral function $a(t,s)=\gR(t,s)-\gA(t,s)$.
The GKB ansatz is defined 
for a more general form of the spectral function,
which implies a corresponding renormalization 
of the free part of the Lagrangian.
Even using more generic form of the spectral function,
our method is applicable 
if the conditions (\ref{gRgA})-(\ref{gAC}) are held
with the replacement of the free-particle spectral function 
$\re^{-\ri\omega(t-s)}$ by renormalized one $a(t,s)$.
(These conditions are nothing but the semi-group property 
 discussed in \Ref{TH}.)
Other parts of the proof 
are based on the retarded or advanced character of the propagators 
which is not affected by the use of generic spectral function.
Thus,
even in the generic GKB formalism,
where the diagrammatic rule may be different due to the renormalization,
the instantaneous-2PR configuration can be eliminated 
if the semi-group property is held for the GKB ansatz.

Note that our method is not valid 
for the time correlation function of $\hat{Q}$ 
such as $\langle \hat{Q}(t)\hat{Q}(s) \rangle$
because we have used the condition that
the external point expressing $Q(\psiC)$ appears 
only once in the diagram.
For the calculation of the time correlation function 
of the composite operator,
some of the instantaneous-2PR configuration may not be canceled.

\section*{Acknowledgements}
The author is very grateful to Prof.~R.~Fukuda for helpful comments.

\end{document}